\documentclass[letter]{aa}

\usepackage{graphicx}
\usepackage{hyperref}
\usepackage{txfonts}
\begin{document} 

   \title{Stellar Halos of Bright Central Galaxies}
   \subtitle{A View from the FEGA Semi-Analytic Model of Galaxy Formation and VEGAS Survey}

   \author{Emanuele Contini \inst{1} 
   \thanks{\email{emanuele.contini82@gmail.com}}
          \and Marilena Spavone  \inst{2}
          \and Rossella Ragusa \inst{2}
          \and Enrichetta Iodice \inst{2}
          \and Sukyoung K. Yi \inst{1}
          }
           \institute{Department of Astronomy and Yonsei University Observatory, Yonsei University, 50 Yonsei-ro, Seodaemun-gu, Seoul 03722, Republic of Korea
            \and
          INAF Osservatorio Astronomico di Capodimonte, Salita Moiariello 16, 80131 Napoli, Italy}

\abstract
{}
{We present theoretical predictions and extrapolations from observed data of the stellar halos surrounding central group/cluster galaxies
and the transition radius between them and the intracluster or diffuse light.}
{Leveraging the state-of-the-art semi-analytic model of galaxy formation, {\small FEGA} (\citealt{contini2024c}), applied to two
dark matter-only cosmological simulations, we derive both the stellar halo mass and its radius. Using theoretical assumptions
about the diffuse light distribution and halo concentration, we extrapolate the same information for observed data from the {\small VEGAS}
survey (\citealt{capaccioli2015,iodice2021}).}
{Our model, supported by observational data and independent simulation results, predicts an increasing transition radius with halo mass, a constant
stellar halo-to-intracluster light ratio, and a stable stellar halo mass fraction with increasing halo mass. Specifically, we find that
the transition radius between the stellar halo and the diffuse light ranges from 20 to 250 kpc, from Milky Way-like halos to large clusters,
while the stellar halo mass comprises only a small fraction, between 7\% and 18\%, of the total stellar mass within the virial radius.}
{These results support the idea that the stellar halo can be viewed as a transition region between the stars bound to the galaxy and those
belonging to the intracluster light, consistent with recent observations and theoretical predictions.}


   \maketitle
%

\section{Introduction}\label{sec:intro}

Stellar halos are extended, diffuse regions of stars that surround the central parts of galaxies, such as the bulge and disk (for a review, see \citealt{arnaboldi2022}). Composed mainly of old, metal-poor stars, these halos provide crucial insights into the formation history of galaxies (\citealt{helmi2008, iodice2016, helmi2018}). The stars in stellar halos are believed to originate from ancient dwarf galaxies that were accreted by the host galaxy through gravitational interactions (\citealt{helmi1999, bullock2005, lane2022, duc2015}), or from early star formation during the galaxy's initial phases of evolution (\citealt{helmi2008}). Unlike stars in the galactic disk, which follow ordered, circular orbits, stellar halo stars are distributed in random, highly elliptical trajectories (\citealt{helmi2020}). The extent of stellar halos can reach up to several hundred kpc in the most massive galaxies, providing a glimpse into past mergers and accretion events that shaped their evolution (\citealt{helmi2018}).

The stars coming from dwarf galaxies are the dominant channel for the formation of stellar halos (e.g., \citealt{cooper2010, harmsen2017, beltrand2024}), but theoretical studies have also shown that some stars were actually formed in situ (\citealt{abadi2006, cooper2015, wright2024}). These in situ stars typically exhibit higher metallicity and are predominantly distributed along the major axis of the galactic disk (\citealt{pillepich2015, monachesi2019}).

Stellar halos exhibit unique properties that distinguish them from other galactic components, i.e., disk and bulge. One key characteristic is their metallicity: stars in these halos are typically metal-poor (\citealt{hartke2022}), indicating that they formed early in the universe, before the interstellar medium became enriched with heavier elements from successive generations of star formation. \cite{spavone2022}, combining data from the Fornax Deep Survey and observations from the Multi Unit Spectroscopic Explorer (MUSE), also investigated the metallicity gradients of stellar halos and their connection to the environment, finding milder gradients in denser environments (see also \citealt{spavone2020,spavone2021}). In terms of age, the stars in stellar halos are generally old, often exceeding 10 Gyr. This supports the theory that these stars are either remnants of early, primitive star formation episodes or were accreted from dwarf galaxies, as previously mentioned.

Stellar halos, being largely the product of accreted stars from stripped or disrupted galaxies that merged in a hierarchical fashion, can exhibit considerable variability in their properties, often in a stochastic manner (\citealt{amorisco2017}). Two clear examples of this are the stellar halos of our own Milky Way and its nearest neighbor, Andromeda. The Milky Way's stellar halo follows a broken power-law density profile (\citealt{deason2014}), whereas Andromeda's stellar halo displays a shallower density profile extending over several kpc (e.g., \citealt{ibata2014}). The stochastic nature of stellar halo formation may significantly influence properties such as colors and metallicity, potentially linking them to the formation of diffuse light, another component that is also thought to form stochastically (\citealt{contini2023}).

Despite being extensively studied theoretically through numerical simulations (references above and therein), stellar halos have often been considered either part of the galaxy or of the diffuse light (e.g., \citealt{contini2014}). In semi-analytic models, the first attempts to model their formation date back to almost two decades ago (e.g., \citealt{bullock2005,delucia2008,cooper2010}). In this Letter, we present a new implementation of stellar halos within a state-of-the-art semi-analytic model (\citealt{contini2024c}), where they are treated as a transition region between the bound stars in the galaxy and the unbound stars of the diffuse light (\citealt{longobardi2015}). To maintain consistency with the properties of stars observed in stellar halos, we assume they primarily originate from the stripping of galaxies during the early stages of diffuse light formation. However, they are formally distinct from the diffuse light stars, as indicated by observations (e.g., \citealt{dsouza2014, longobardi2015}) when using spectroscopy, but the separation between the two components is not trivial when looking at the light profiles.

In Section \ref{sec:methods}, we detail the implementation of stellar halo formation in our model. In Section \ref{sec:results}, we present our analysis, showing that the transition radius between the diffuse light and stellar halos, defined as the typical radius of the latter, aligns well with independent measurements from simulations (\citealt{proctor2024}).  Finally, in Section \ref{sec:conclusions}, we discuss our findings and highlight the key conclusions.

Throughout the rest of the paper, stellar and halo masses are corrected for $h = 0.68$, and the stellar masses are derived under the assumption of a Chabrier initial mass function (\citealt{chabrier2003}).

\section{Methods}\label{sec:methods}

The semi-analytic model {\small FEGA} (Formation and Evolution of GAlaxies) was detailed in \cite{contini2024c}, a state-of-the-art model that incorporates the latest prescriptions for the baryonic physics involved in galaxy formation, including a revised star formation law and a novel implementation of the positive feedback from active galactic nuclei. For a comprehensive discussion of the model, we refer the reader to that paper. Here, we focus on the new features introduced in this version, specifically the formation of stellar halos (SHs) and the definition of the transition radius between SHs and the surrounding intragroup/cluster light (IGL/ICL) or diffuse light \footnote{In this Letter, we use the terms ICL and diffuse light interchangeably.}.

Before delving into the specifics of the SH implementation, we summarize the primary mechanisms of diffuse light formation in {\small FEGA}. Diffuse light can form through three main channels: stellar stripping of satellite galaxies, mergers, and pre-processing. In {\small FEGA}, a large portion of diffuse light arises from stars stripped as satellites orbit within the host halo's potential well, particularly in the densest regions, where tidal interactions are strongest (stellar stripping channel). Another significant contribution comes from mergers between the central galaxy (CG) and satellites, where violent relaxation processes redistribute part of the stellar mass into the diffuse component (merger channel). Additionally, diffuse light can form in groups and later be accreted by larger structures during assembly (pre-processing channel), though this is essentially a sub-channel, as it originates from either stellar stripping or mergers. For more details on the specific prescriptions of each channel, we refer the reader to \cite{contini2024c}.

We assume the SH to form from stars belonging to the ICL that are very close to the CG. In order to do it, we need to define a radius within which stars in the SH are distributed, a radius that hereafter we call transition radius, $R_{\rm{trans}}$. Following \cite{contini2020b}, we assume that, at each time, the ICL follows an NFW profile (\citealt{nfw1997}) with a higher concentration than the dark matter (DM) halo in which it resides (\citealt{harris2017,pillepich2018,montes2019,contini2022}). Specifically, under this assumption, the ICL concentration is defined as:
\begin{equation}\label{eqn:conc}
 c_{\rm{ICL}} = \gamma c_{200} = \gamma \frac{R_{\rm{200}}}{R_{\rm{s, DM}}} = \frac{R_{\rm{200}}}{R_{\rm{s, ICL}}} \, ,
\end{equation}
where $R_{\rm{200}}$ and $R_{\rm{s, DM}}$ represent the virial and scale radii of the DM halo, respectively, and $R_{\rm{s, ICL}}$ the scale radius of the ICL distribution. As in \cite{contini2020b}, the parameter $\gamma$ is assumed to be greater than 1 to reflect a more concentrated ICL distribution compared to the DM. From Equation \ref{eqn:conc}, the scale radius of the ICL distribution can be derived as:
\begin{equation}\label{eqn:rtrans}
  R_{\rm{s, ICL}} = \frac{R_{\rm{200}}}{c_{\rm{ICL}}} \, .
\end{equation}
We assume that $R_{\rm{trans}}=R_{\rm{s, ICL}}$, and all the stars within this radius, at any given time, belong to the SH. This new component, composed of stars coming from the ICL, is considered as a transition between the bound stars in the galaxy and the unbound stars in the ICL. However, under certain conditions, these stars can be accreted into the galaxy. Specifically, if $R_{\rm{trans}}$ is smaller than the size of the galaxy's bulge, all the stars within it, belonging to the SH, are moved into this component. Additionally, a constraint is placed on the SH mass: for dynamical stability, the semi-analytic model does not allow the SH to exceed the total mass of the galaxy, and when this occurs, the excess stars are transferred to the galaxy’s disk component.

It is important to note that a semi-analytic framework does not inherently provide information about whether stars within the SH component are gravitationally bound to the CG or not; rather, this is left to the user’s interpretation. As mentioned in Section \ref{sec:intro}, observations typically consider SH stars to be bound to the CG, forming a third stellar component within it. However, in our implementation, the model’s flexibility allows the SH to be considered either as a third component of the CG (and thus bound to it) or as a subset of the diffuse light (and therefore unbound to the CG).

In \cite{contini2020b}, the parameter $\gamma$ was assumed to be 3 to align with the observed CG+ICL mass within 100 kpc, as noted by \cite{demaio2020} at different redshifts. In this study, we followed the same approach as in \cite{contini2024c} for model calibration, allowing an MCMC algorithm to determine the range of values that best reproduce the observations. Consequently, we first ran {\small FEGA} on partitions of the merger trees from the dark matter-only cosmological simulations YS50HR and YS200 (details in \citealt{contini2023,contini2024}) to constrain $\gamma$, and then applied it to the full merger trees to construct the galaxy catalogs. Our calibration suggests that $\gamma$ varies between 1 and $\sim3$, so the model for the final catalogs assumes a random value for each ICL component within this range.

To compare the model predictions with observational data, we utilize the measurements of the ICL in galaxy groups and clusters provided by \cite{ragusa2023} from the VST Early-type GAlaxy Survey ({\small VEGAS}, \citealt{capaccioli2015,iodice2021}, and the data analysis is extensively described in \citealt{ragusa2021,ragusa2022}). The sample consists of 22 objects, covering the core regions of groups and clusters in the local universe at $z<0.05$, and in the halo mass range $[0.4-26.9]\cdot 10^{13} \rm{M_{\odot}}$. In this Letter, we focus on the 17 objects detailed in \cite{ragusa2023}. {\small VEGAS} is a multi-band, deep imaging survey based on observations acquired with the ESO VLT Survey Telescope (for details, see \citealt{schipani2012}). Data reduction was performed using the AstroWISE pipeline (details in \citealt{venhola2017} and references therein) and the VST-Tube pipeline (\citealt{capaccioli2015}), which provide comparable results. For a comprehensive description of the image data reduction, we refer the reader to \cite{capaccioli2015}, \cite{iodice2016} and \cite{spavone2017}. All 17 objects in our sample were observed in the \emph{g} and \emph{r} bands, with some also observed in the \emph{u} and/or \emph{i} bands.

These data encompass all the diffuse light, but we need to extrapolate the SH, which we assume to be a transition region. To address this issue, we apply the same assumption used in the model, specifically an NFW profile for the ICL distribution in order to derive $R_{\rm{trans}}$. Unfortunately, the halo concentration is not a directly observable quantity, yet it is essential for defining $R_{\rm{trans}}$. Therefore, we utilize the halo concentration-mass relation (e.g., \citealt{prada2012,correa2015,child2018}) from \cite{correa2015} to derive $c_{200}$ from the virial mass and, using a random $\gamma$ within the same range mentioned above, we calculate $c_{\rm{ICL}}$ from Equation \ref{eqn:conc} first, and finally $R_{\rm{trans}}$ from Equation \ref{eqn:rtrans}. With this information and the assumption of an NFW profile, we can isolate the light in the transition region, which we consider to be the SH in this Letter, from the ICL measurements reported by \cite{ragusa2023}.

\section{Results}\label{sec:results}

\begin{figure}
\centering
\includegraphics[width=0.48\textwidth]{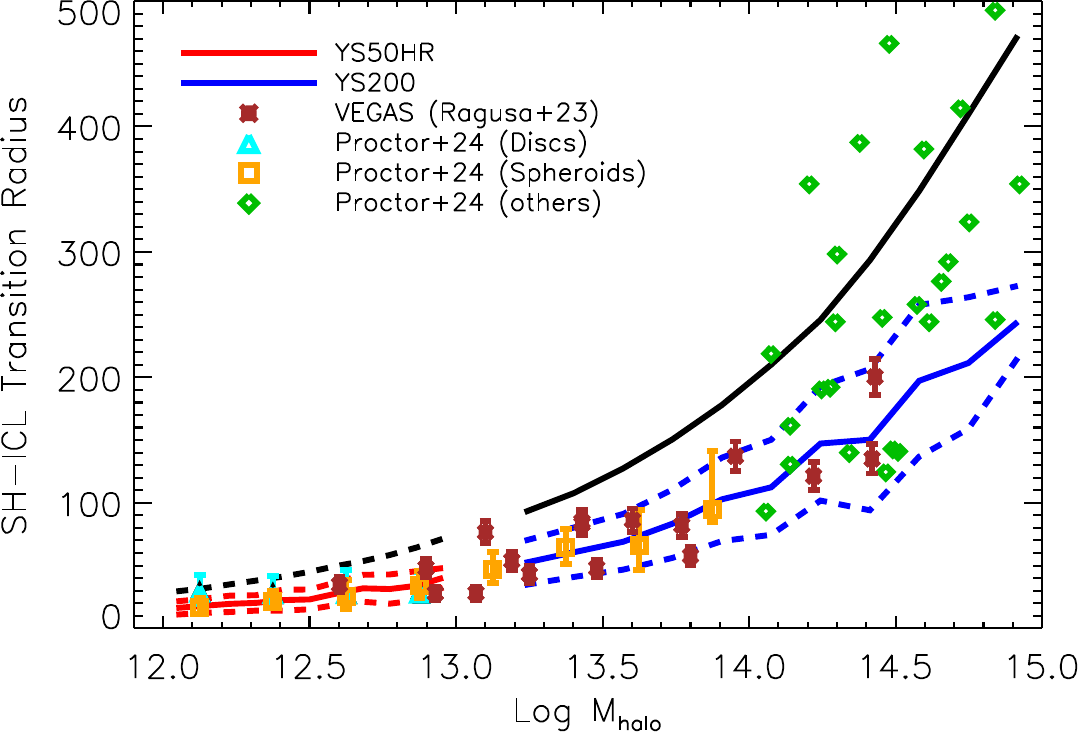}
\caption{Transition radius between the SH and the surrounding ICL as a function of halo mass. Our model predictions (red and blue lines) are compared with the results from \cite{proctor2024} based on the Eagle (cyan and orange symbols) and C-Eagle (green symbols) simulations, and extrapolations of observed data by \cite{ragusa2023} from the {\small VEGAS} survey (brown symbols). For comparison, we show with dashed (YS50HR) and solid (YS200) black lines the scale radius-halo mass relation. Our model predictions are in very good agreement with both theoretical and observational measurements across the entire halo mass range investigated, and all data indicate an increasing transition radius with halo mass.}
\label{fig:rtrans}
\end{figure}

We now proceed with the analysis, starting by plotting, in Figure \ref{fig:rtrans}, the transition radius, $R_{\rm{trans}}$, as a function of the halo mass of CGs. Overlaid on our model predictions (red and blue lines \footnote{The gap in halo mass between the two predictions is given by the box of the simulations. In YS50HR we avoid halos more massive than $\log M_{\rm{halo}}=13$ in order to have enough statistics, and halos less massive than $\log M_{\rm{halo}}\sim 13.2$ in YS200 for reasons linked to the resolution.}) are the independent measurements of the transition radius by \cite{proctor2024} from the Eagle (\citealt{schaye2015,crain2015}) and C-Eagle (\citealt{bahe2017,barnes2017}) simulations (cyan, orange, and green symbols), along with our extrapolations of the observed data by \cite{ragusa2023} from the {\small VEGAS} survey (brown symbols). Overplotted by dashed (YS50HR) and solid (YS200) lines, we show the scale radius-halo mass relation.

Regarding $R_{\rm{trans}}$ by Proctor et al. 2024, its definition is fundamentally different from ours. Specifically, to compute the transition radius, Proctor et al. 2024 constructed spherically averaged density profiles for each component of the galaxy and then interpolated these profiles to determine the radius at which the density of the ICL exceeded that of the other components.

From Figure \ref{fig:rtrans}, we observe that $R_{\rm{trans}}$ increases with halo mass, ranging from a few tens of kpc in Milky Way-like halos to 250 kpc in massive clusters. Importantly, our predictions align excellently with the measurements by Proctor et al. 2024, as well as with our extrapolations from the {\small VEGAS} data.

\begin{figure}
\centering
\includegraphics[width=0.48\textwidth]{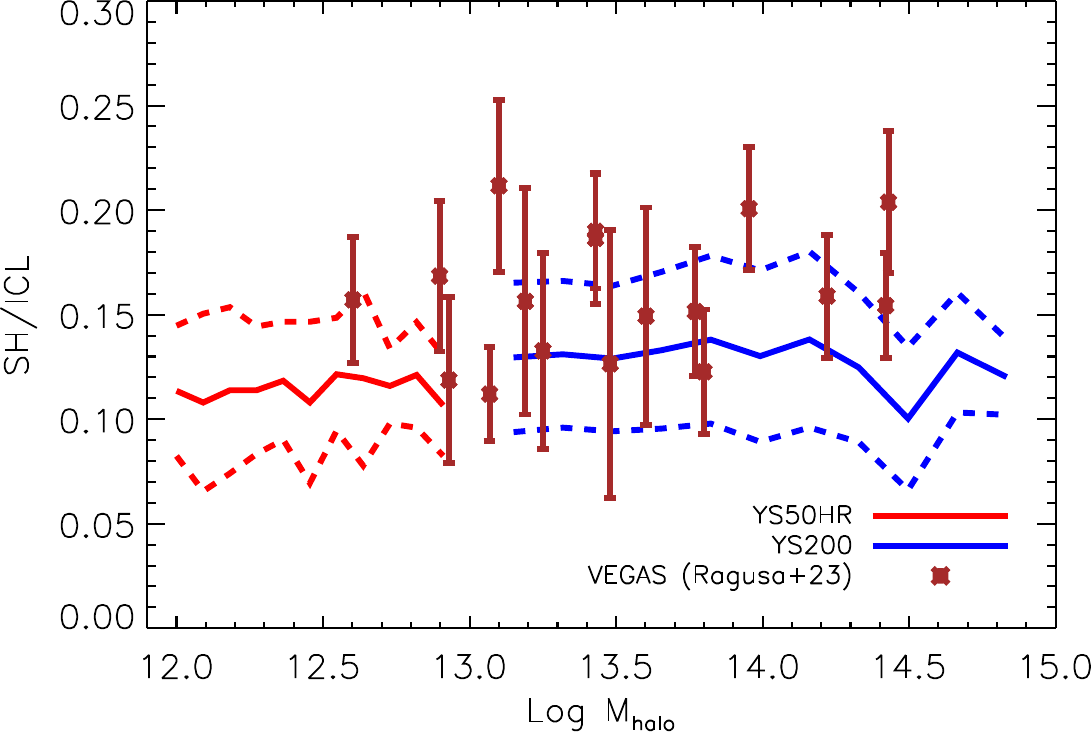}
\caption{Ratio between the SH-ICL mass as a function of halo mass. The predictions from our model (red and blue lines) are compared to the extrapolated observed data by \cite{ragusa2023} from the {\small VEGAS} survey (brown symbols). Our model aligns reasonably well with the observations, and both indicate no significant relation with halo mass.}
\label{fig:shicl}
\end{figure}

In Figure \ref{fig:shicl}, we examine the mass fraction between the SH and the ICL as a function of halo mass. This relationship provides insight into the significance of the SH compared to the more extended ICL. Our model predictions are compared with extrapolations from the {\small VEGAS} data. Both the model and the data indicate no correlation with halo mass, with the SH mass being between 7\% and 18\% (scatter included) of the ICL mass, and most of the {\small VEGAS} data support this by remaining within this range. This suggests that, at present, up to one-fifth of the ICL formed during the assembly of the halo has transitioned to the SH region. An interesting aspect to explore further would be the evolution of this relationship with redshift, which we will address in a separate study.

\begin{figure}
\centering
\includegraphics[width=0.48\textwidth]{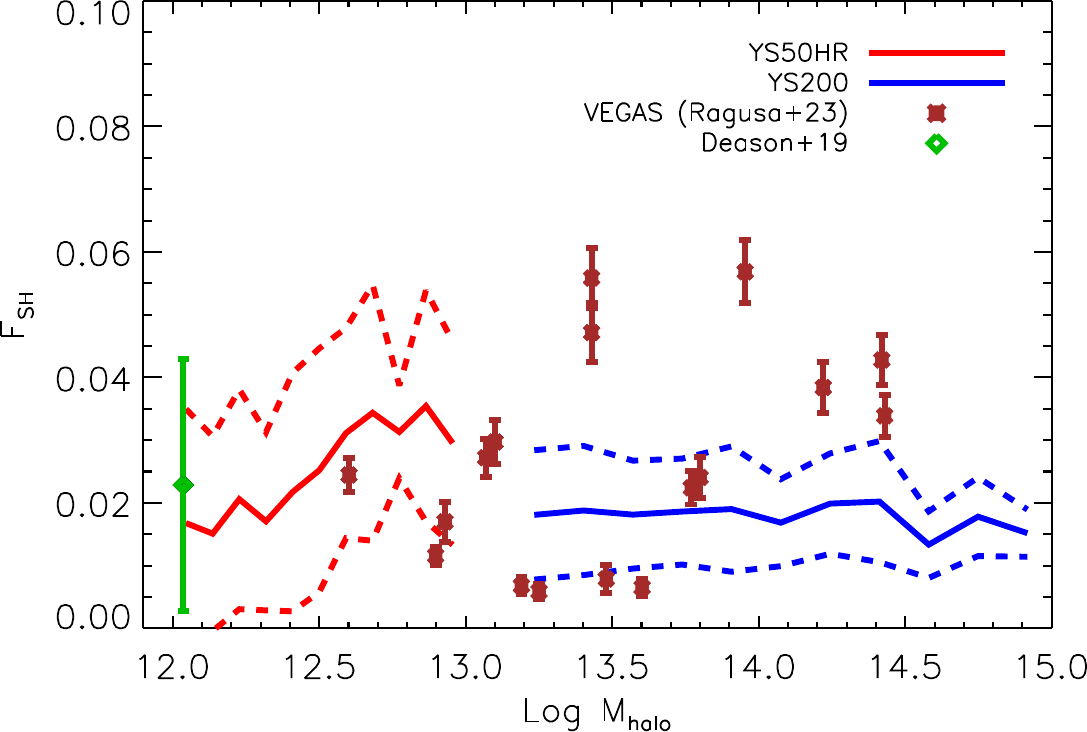}
\caption{Fraction of SH mass with respect to the total mass within the virial radius of the halo presented as a function of halo mass. Similar to the ICL fraction in the same halo mass range (\citealt{contini2024b}), there is no correlation observed in groups and clusters. However, there are indications of a decreasing fraction on smaller scales than loose groups, once the resolution of the simulations is taken into account. Our model predictions (red and blue lines) are compared with the extrapolated observed data by \cite{ragusa2023} from the {\small VEGAS} survey (brown symbols) and the SH mass fraction reported by \cite{deason2019} for the Milky Way (green symbol).}
\label{fig:fSH}
\end{figure}

To conclude our analysis, in Figure \ref{fig:fSH}, we present the fraction of stellar mass in the SH relative to the total stellar mass within the virial radius of the halo as a function of halo mass. As with the previous figures, the model predictions are compared with extrapolations from the {\small VEGAS} survey data, along with the measurements by \cite{deason2019} for our galaxy. There is no correlation observed across almost the entire halo mass range probed, although there is a hint of a decreasing fraction at smaller scales predicted by YS50HR, which, once the resolution of the simulation is taken into account, should align with YS200. This finding is a direct consequence of the lack of correlation seen in Figure \ref{fig:shicl}, in conjunction with the known relationship between the ICL fraction and halo mass (e.g., \citealt{contini2024b}). The {\small VEGAS} data show reasonable agreement at low mass scales but tend to be higher for more massive objects.

In light of these results, we now proceed to the next section, where we discuss our findings and draw conclusions.

\section{Discussion and conclusions}\label{sec:conclusions}

The typical radius at which the ICL begins to dominate the light distribution has been investigated both observationally and theoretically by several authors (\citealt{iodice2016,spavone2020,gonzalez2021,montes2021,contini2022,proctor2024,spavone2024} and references therein) in recent years. From an observational perspective, the transition radius $R_{\rm{trans}}$, which we remind the reader represents the point where ICL light predominates in the distribution—i.e., where the CG+ICL light distribution exhibits a clear break—has been found to be approximately 60–80 kpc. These values have also been confirmed by \cite{contini2022}, where $R_{\rm{trans}}$ was defined as the typical radius at which the ICL mass comprises given percentages ($>50$\%) of the total mass locally. In that study, the authors employed the same assumptions used here to determine the scale radius of the ICL distribution, although the parameter $\gamma$ in this analysis is better constrained.

Nevertheless, despite the good agreement between observations and theory, \cite{contini2022} did not formally consider or define any transition region between the CG and the ICL, in contrast to the present work. Indeed, their definition of $R_{\rm{trans}}$ relies solely on the mass distribution of the CG and ICL. Here, we define $R_{\rm{trans}}$ as the radius that characterizes the SH, which in turn defines the transition region between the CG and ICL. An independent study by \cite{proctor2024} employed a similar (but not identical) definition of $R_{\rm{trans}}$ to that used in \cite{contini2022}. While the latter focused on the mass profiles of the CG and ICL, Proctor et al. 2024 considered the density profile. They clearly demonstrated that, due to their definition, $R_{\rm{trans}}$ is an increasing function of halo mass. As shown in Figure \ref{fig:rtrans}, this is also true for our predictions, which are more closely linked to the results of Proctor et al. 2024 than to those of \cite{contini2022}\footnote{Here, $R_{\rm{trans}}$ is derived directly from the definition of halo concentration, which depends on the virial radius and the scale radius. The computation of the latter is based on the density profile.}, but remains independent of them.

An increasing $R_{\rm{trans}}$ with halo mass is also more physically understandable. Indeed, more massive CGs reside in more massive halos, which also have a larger amount of ICL due to the larger number of galaxies that are subject to stripping or that can merge. From this perspective, it is reasonable to expect a more extended transition region between the two components, i.e., a larger SH. A key point of this Letter is that our new definition of the transition radius not only agrees very well with independent predictions from simulations, but it is also capable of isolating a region where stars do not formally belong to the galaxy's bulge or disk, nor do they belong to the ICL anymore. We have assumed this region to be the stellar halo that is observed in nearby galaxies and well measured in our own.

To demonstrate the validity of our definition, and thus our assumptions, the predicted SH mass must be comparable to observed values. In Figure \ref{fig:shicl}, we first compared the ratio between the mass in the SH and that of the ICL. Considering the scarcity of SH measurements outside the local group (e.g., \citealt{beltrand2024}), given the necessity of resolving the stellar kinematics, we used the same assumptions employed in our model to extrapolate the SH from the {\small VEGAS} ICL measurements. These data suggest that the mass in SHs can range between 5\% and 20\% (including scatter) of that in the ICL, which aligns very well with our model predictions, ranging from 7\% to 18\%.

Similarly, in Figure \ref{fig:fSH}, we showed that our model predicts a fraction of SH mass with respect to the total stellar mass of up to around 5\%, indicating that the SH is less significant than both the bulge and disk of the galaxy combined, as well as the ICL. According to recent results by \cite{brown2024}, who investigated the assembly of the ICL in 10 groups and clusters extracted from the {\small Horizon-AGN} simulation, the SH appears to be the least important component, following satellite galaxies, CGs, and ICL in terms of stellar mass within the virial radius. However, there is no trend with halo mass in either case, regarding the ICL mass and the total stellar mass. Considering that the ICL fraction also does not depend on halo mass at group and cluster scales, this implies that more extended SHs are hosted by more massive halos (Figure \ref{fig:rtrans}), but their larger mass is balanced by the greater mass in ICL and total stellar mass (Figures \ref{fig:shicl} and \ref{fig:fSH}).

Clearly, our analysis is not without uncertainties related to how we treated the observed data. The information we used to extrapolate our results comes from separating the CG and the ICL, a process that can lead to underestimating or overestimating the ICL depending on the masking applied to nearby satellite galaxies and the method used for the separation. This may account for the larger scatter observed in the {\small VEGAS} data.

To conclude, based on the analysis presented in Section \ref{sec:results} and the considerations discussed above, we highlight the key points and conclusions of this Letter:
\begin{itemize}
 \item The SH can be associated with the transition region between the galaxy and the ICL, and its radius, $R_{\rm{trans}}$, can be regarded as the boundary between the CG and the ICL, through the SH;
 \item Our new definition of $R_{\rm{trans}}$ aligns well with that used in \cite{proctor2024}. In both cases, the transition radius increases with halo mass (Figure \ref{fig:rtrans}), indicating that SHs in clusters are more extensive than those in smaller groups;
 \item The fraction of SH mass relative to the ICL and total stellar mass (Figures \ref{fig:shicl} and \ref{fig:fSH}) is in good agreement with extrapolations from observed data (from the {\small VEGAS} survey) and with measurements by \cite{deason2019} for the stellar halo of our galaxy. Generally, the SH represents only a small fraction of the total stellar mass within the virial radius of the halo, typically around 7\% to 18\% (model predictions) or 5\% to 20\% (extrapolated from observed data) of the ICL mass.
\end{itemize}

The next step in investigating SH formation is to repeat the analysis as a function of redshift and examine potential dependencies on halo formation time and its dynamical state, which is likely considering recent findings such as those by \citealt{ragusa2023}, \citealt{contini2023}, \citealt{contini2024b}, and others for the ICL. Furthermore, we aim to study how properties like the colors and metallicity of the SH evolve over time and explore their potential connection with those of the ICL.


\begin{acknowledgements}
The authors thank the anonymous referee for their constructive comments, which have significantly contributed to improving this manuscript.
E.C. and S.K.Y. acknowledge support from the Korean National Research Foundation (2020R1A2C3003769), (2022R1A6A1A03053472), and E.C. acknowledges support from the Korean National Research Foundation (RS-2023-00241934). M.S. and E.I. acknowledge the support by the Italian Ministry for 1224 Education University and Research (MIUR) grant PRIN 2022 2022383WFT 1225
“SUNRISE”, CUP C53D23000850006 and by VST funds. R.R. acknowledges financial support through grants PRIN-MIUR 2020SKSTHZ and through INAF-WEAVE StePS founds.

\end{acknowledgements}

 \bibliographystyle{aa.bst}
  \bibliography{biblio}

\end{document}